# Phase conjugated master oscillator fiber power amplifier

**Tingwei Gu[1,2], Xin Zeng[1], Huawei Jiang[1], Suming Luo[1,2], Maokai Yang[1], Xuezong Yang[1], Yan Feng[1,*]**

1.Hangzhou Institute for Advanced Study, University of Chinese Academy of Sciences, Hangzhou 310024, China

2.Shanghai Institute of Optics and Fine Mechanics, Chinese Academy of Sciences, Shanghai 201800, China

*yfeng@ucas.ac.cn

**Abtract：**

High-power narrow-linewidth fiber lasers are fundamentally limited by stimulated Brillouin scattering (SBS), which constrains further power scaling while maintaining spectral linewidth. Traditional mitigation techniques, such as active phase modulation, often introduce trade-offs among complexity, cost, and spectral brightness. In this study, we propose and experimentally demonstrate a novel all-optical approach for spectral linewidth manipulation and SBS suppression using optical phase conjugation (OPC). By leveraging nonlinear spectral broadening followed by phase conjugation, this method enables sophisticated linewidth narrowing in fiber amplifier, resulting in narrow linewidth output and enhanced SBS threshold. Using a low-cost fiber oscillator as the seed source, we achieve a spectral compression ratio exceeding 3 times. This method not only eliminates the need for complex electro-optic modulation systems but also provides a pathway toward simpler, high-brightness fiber laser systems. Our findings underscore the viability of OPC as a transformative tool for nonlinearity management and power scaling in high-performance fiber laser architectures.

**Introduction**

Over the past decades, fiber laser technology has emerged as a dominant force in the laser industry, distinguished by its high efficiency, compactness, excellent beam quality and outstanding versatility[1–4]. A central pursuit in the field remains power scaling—yet fundamental constraints impede further progress. Both theoretical and experimental studies confirm that the output power of diffraction-limited fiber lasers is limited to several tens of kilowatts, constrained by intrinsic nonlinear effects, thermal load, and fiber mechanical reliability[5–7]. To transcend these barriers, beam-combining techniques such as coherent beam combining (CBC)[8] and spectral beam combining (SBC)[9] have been developed, enabling power scaling without sacrificing beam quality. These approaches, however, impose stringent spectral requirements on individual emitter linewidths. In CBC, extended coherence is essential for stable interference, while in SBC, finite linewidth introduces angular dispersion that degrades beam quality[10]. For Yb-doped fiber lasers at 1 μm, achieving high combining quality mandates linewidths below 0.1 nm[9,11]. Consequently, high-power narrow-linewidth fiber lasers have become indispensable.

Nevertheless, power scaling in such systems confronts several nonlinear effects, including SBS, transverse mode instability, and stimulated Raman scattering. Among these, SBS represents the principal limiting factor for narrow-linewidth fiber laser, with a threshold power intimately tied to laser linewidth. To raise the SBS threshold without compromising spectral linewidth, diverse suppression strategies have been explored. These include specialty fibers with reduced Brillouin gain coefficient, strain or thermal gradient engineering that decreases effective Brillouin gain, and innovative amplifier designs that curtail effective interaction lengths[12–14]. Alternatively, active linewidth broadening via phase modulation—using sinusoidal, white-noise, pseudo-random binary sequences, or optimized waveforms[15–21]— has been employed. Yet these techniques face practical limitations in cost, complexity, and scalability.

A promising alternative lies in methods that cyclically broaden and recompress the laser linewidth. One strategy employs electro-optic phase modulation to single frequency seed laser followed by active demodulation after fiber amplifier, though limited power-handling capabilities and bandwidth restrict its utility beyond hundred-watt level[22,23]. Passive demodulation via nonlinear frequency conversion has also been demonstrated, albeit with trade-offs in wavelength shift and conversion efficiency[24,25]. Notably, Goodno *et al.* introduced a co-modulation technique combining phase and amplitude modulation, leveraging self-phase modulation (SPM) in fibers to achieve passive spectral compression[26,27]. This approach enhances SBS suppression by reducing integrated nonlinear gain relative to conventional phase-modulated master oscillator fiber amplifiers at identical output linewidth. However, its practical adoption is hampered by system complexity and electronic bandwidth limitations.

While Kerr nonlinearity typically induces spectral broadening during amplification, proper pre-shaping of the spectral phase can conversely lead to spectral narrowing. This insight underscores the importance of sophisticated phase management in seed lasers for high-power narrow-linewidth laser system. In this work, we introduce a novel approach based on optical phase conjugation (OPC) for spectral compression and SBS suppression. Unlike electro-optic modulation schemes, our technique exploits purely optical processes: initial nonlinear spectral broadening, followed by optical phase conjugation, which inverts the nonlinear phase. The conjugated nonlinear phase can be cancelled by subsequent nonlinear phase accumulated in the fiber amplifier with proper setting, effectively recovering the original linewidth. In the experiment, we demonstrate about 3.15 times spectral compression and about 2.5 times enhancement in SBS threshold. With its all-optical high-bandwidth nature, this method offers compelling advantages in scalability and flexibility over conventional modulation techniques. This paradigm may open new avenues for power scaling in narrow-linewidth fiber lasers and reshape nonlinear management strategies in high-power laser systems.

**Results**

**Principle of the approach**

Figure 1a presents a schematic of the proposed approach. A narrow-linewidth seed laser is first broadened spectrally through nonlinear optical effects. OPC is then applied to the broadened signal before it is injected into the fiber amplifier. Under appropriate parameter settings, the laser system undergoes a cycle of spectral broadening followed by compression, ultimately yielding a narrow-linewidth output at high power. Since the linewidth remains broader than the final output throughout the amplification stage—and considering that the SBS gain coefficient is approximately inversely proportional to linewidth—the integrated SBS gain is reduced compared to systems with constant linewidth, such as those employing conventional phase modulation. It is therefore evident that OPC serves as the cornerstone of this method: by inverting the spectral phase, it enables nonlinear phase demodulation during subsequent amplification.

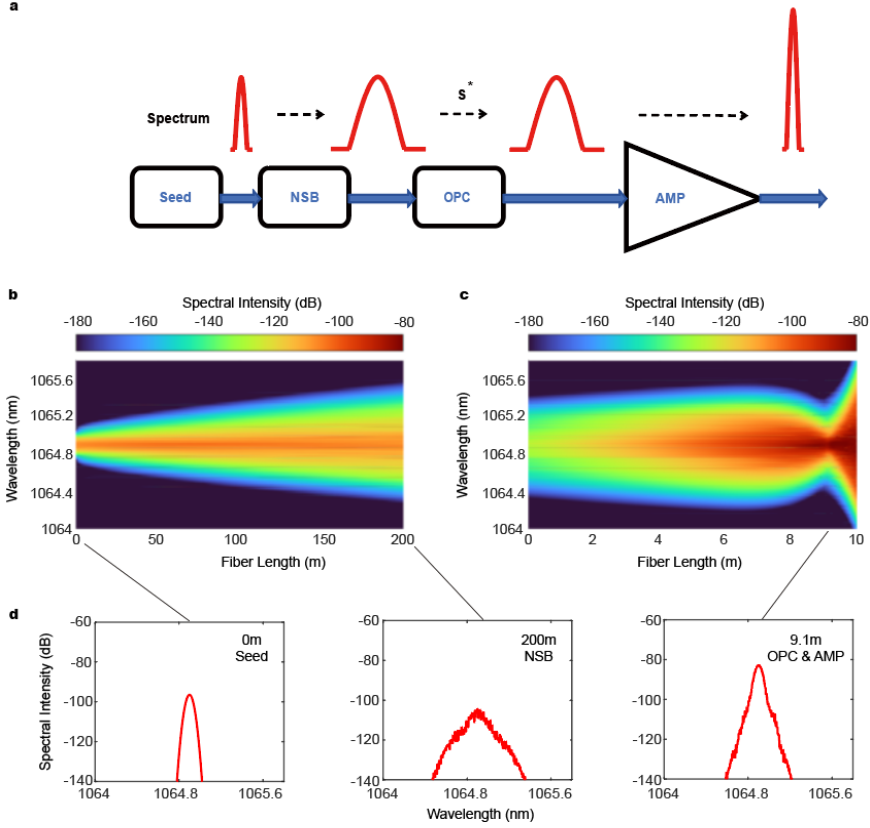

Fig. 1 **Spectral evolution of a phase conjugated master oscillator fiber power amplifier. a**, schematic of the proposed approach. **b**, simulated spectral broadening in a nonlinear optical fiber. **c**, simulated spectral evolution in an amplifier after OPC. **d**, Spectrum of initial seed (0m), after nonlinear spectral broadening (NSB) (200m), at optimized compression position (9.1m in the amplifier), respectively.

The propagation of light in a nonlinear dispersive medium such as optical fiber is governed by the nonlinear Schrödinger equation (NLSE)[28]:

$$\frac{\partial A}{\partial z} + \frac{i}{2}\beta_2 \frac{\partial^2 A}{\partial t^2} = i\gamma |A|^2 A \tag{1}$$

where $A$ represents the complex field amplitude, $\beta_2$ denotes the group-velocity dispersion coefficient, and $\gamma$ is the Kerr nonlinear coefficient. Taking the complex conjugate of both sides yields:

$$\frac{\partial A^*}{\partial (-z)} + \frac{i}{2}\beta_2 \frac{\partial^2 A^*}{\partial t^2} = i\gamma |A^*|^2 A^* \tag{2}$$

A comparison between Equation. (1) and (2) shows that the conjugated field $A^*$ satisfies the same NLSE, albeit with a reversal in the sign of the z-coordinate. This sign change implies that the conjugated wave effectively experiences reversed dispersion and nonlinearity compared to the original wave. Consequently, OPC generates a conjugated equivalent that retraces the original wave's path in reverse. As a result, nonlinear phase accumulation and dispersive effects from the initial propagation are systematically compensated during the conjugate wave's transmission, ultimately restoring the original signal properties.

OPC has been employed in optical communications for signal regeneration via mid-span phase conjugation[29,30]. This application requires precise symmetry in fiber span length and dispersion to mitigate cumulative temporal distortion in systems where dispersive effects dominate over nonlinear contributions. In contrast, high-power fiber laser systems—especially under continuous-wave (CW) operation—are predominantly impaired by nonlinearities, which induce significant spectral broadening. Here, the phase-inverting property of OPC enables effective compensation of nonlinear phase accumulation, thereby restoring narrow-linewidth output without imposing strict symmetry conditions. This fundamental difference in physical mechanisms allows us to strategically harness the linewidth recovery capability in amplifier architectures. By intentionally introducing pre-amplification nonlinear spectral broadening, the SBS threshold is enhanced, while the phase-conjugation process ensures recovery of a narrow spectrum output.

To elucidate the spectral evolution dynamics, we conducted numerical simulations based on the NLSE solved via the split-step Fourier method (SSFM). Representative results are shown in Fig. 1b and 1c, where light transmission through two portions of the laser system (see Methods for details) is simulated. Fig. 1b illustrates the initial spectral broadening of the seed laser due to self-phase modulation in a nonlinear optical fiber. After ideal phase conjugation and attenuation by 20 dB, the signal is amplified in subsequent fiber amplifier, with its spectral evolution depicted in Fig. 1c. Prominent spectral compression occurs at approximately 9.1 m, beyond which the spectrum broadens again with further propagation. This behavior confirms the intended process of nonlinear phase modulation followed by conjugation-assisted demodulation within the amplifier. The optimum compression corresponds to full cancellation of the nonlinear phase shifts accumulated before and after conjugation—a condition quantified by the B-integral, which defines the parameters for maximum spectral narrowing. The relevant expressions are:

(3)

(4)

where $\gamma_1$ and $\gamma_2$ represent the nonlinear coefficients of the initial broadening fiber and the amplifier fiber, respectively; $L_1$ and $L_2$ denote the corresponding fiber lengths; and $\varphi_{N1}$ and $\varphi_{N2}$ are the B-integrals for each segment. Equation. (3) says that the nonlinear phase shift acquired before conjugation should be equal in magnitude to that introduced during amplification. The negative sign in Equation. (4) points the critical role of conjugation in enabling phase cancellation and consequent spectral compression.

Fig. 1d compares the initial seed spectrum, the broadened spectrum after 200 m of propagation with that of the conjugated signal after traversing 9.1 m in the amplifier. The amplified conjugated spectrum exhibits significant energy re-concentration into a narrowband component nearly identical to the original seed, confirming effective phase demodulation. Residual spectral pedestal persists, however, is by more than 20 dB lower to the central peak. These features are attributed to dispersion, gain/loss, and structural asymmetry between the pre- and post-conjugation stages. Further detailed simulations and analysis of these effects are provided in Supplementary Information Section 1, confirming that asymmetry is a primary factor influencing spectral pedestal generation.

**Laser architecture**

Figure 2a illustrates the experimental setup (see Methods for details), and Fig. 2b displays the measured spectral evolution throughout the laser system, which has a B-integral of approximately 6.1 rad. The seed laser is a conventional fiber oscillator based on a pair of fiber Bragg gratings (FBG) centered at 1064.9 nm. Although more complex seeds utilizing phase and amplitude modulation were initially explored, the traditional FBG-based oscillator was found to produce enhanced spectral broadening—a trait disadvantageous in conventional amplification but advantageous here, as it promotes more rapid spectral power redistribution for a given B-integral. After the master oscillator, a pre-amplifier adjusts the seed power entering the subsequent nonlinear fiber to precisely control the amount of nonlinear phase accumulation. The spectra of the original seed (black) and after nonlinear broadening (blue) are shown in Fig. 2b. Notably, the spectral modulation arises solely from nonlinear optical effects, broadening the linewidth from 0.049 nm to 0.205 nm. Full characteristics of the oscillator and nonlinear broadening process are provided in Supplementary Information Section 2.

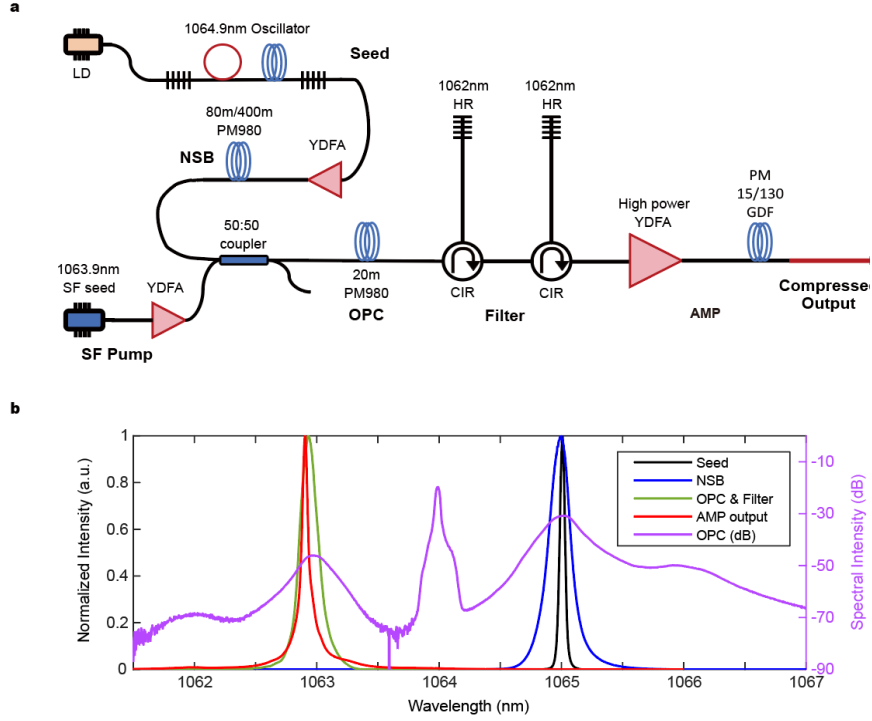

Fig. 2 **Experimental setup and spectral evolution. a**, Schematic of the laser system. A FBG-based oscillator serves as the seed laser, followed by a pre-amplifier and a nonlinear broadening fiber. The OPC stage consists of a single-frequency pump laser, a 50:50 fiber coupler, a 20 m-long PM980 fiber, and two spectral filters implemented using circulators and FBGs. The conjugated signal is amplified in a high-power fiber amplifier with an extended delivery fiber. **b**, Measured optical spectra at different stages: original seed (black), after nonlinear broadening (blue), after OPC (purple), after filtering (green), and after amplification (red). Here the nonlinear phase shift (B-integral) of the laser system is approximately 6.1 rad.

Optical phase conjugation is implemented via degenerate four-wave mixing (FWM) in a standard polarization-maintaining fiber. The broadened seed is combined with a 1063.9 nm single-frequency pump laser using a 50:50 coupler and launched into a 20 m span of PM980 fiber, generating a phase-conjugated signal at approximately 1062.9 nm. The output spectrum from the FWM process is shown in purple in Fig. 2b. Among various OPC generation techniques, the FWM approach offers advantages in fiber compatibility and accessibility[31,32].

Analysis of the OPC process (see in Supplementary Information Section 3) highlights two considerations: first, the conjugated signal acquires twice the pump phase, which cannot be demodulated in subsequent nonlinear propagation—necessitating the use of a single-frequency pump.

Second, phase mismatch in the FWM process reduces conjugation efficiency and introduces a wavelength-dependent phase shift on the conjugated signal, collectively impairing final compression performance.

The phase-conjugated signal at ~1062.9 nm is spectrally filtered (Fig. 2b, green) and injected into a high-power fiber amplifier. The output spectrum (red) shows a compressed linewidth of 0.063 nm, though residual sidebands remain—attributed to system asymmetry and FWM phase mismatch.

Having outlined the spectral evolution of the laser system, the following section will discuss the detailed propagation dynamics of the conjugated signal within the high-power amplifier.

**Nonlinear spectral compression**

The high-power fiber amplifier utilized in this experiment was originally optimized for single-frequency amplification, exhibiting low intrinsic nonlinearity. To induce sufficient nonlinear spectral manipulation, a long delivery fiber was intentionally spliced after the amplifier to enhance the nonlinear phase accumulation. Fig. 3a shows the output spectral evolution at different power levels for a system incorporating a 50 m PM-15/130-GDF amplifier delivery fiber and a 400 m PM980 nonlinear broadening fiber prior to OPC. At an output power of 105.4 W, the estimated B-integral is approximately 6.1 rad. As visible in Fig. 3a, imperfect filtering following OPC resulted in residual spectral components: the second-order FWM product at ~1061.9 nm and the remnant pump laser at 1063.9 nm both remain present. These filtering imperfections significantly alter spectral evolution and contribute to prominent sideband generation. Fig. 3b presents corresponding simulation results that closely reproduce the experimental spectral behavior.

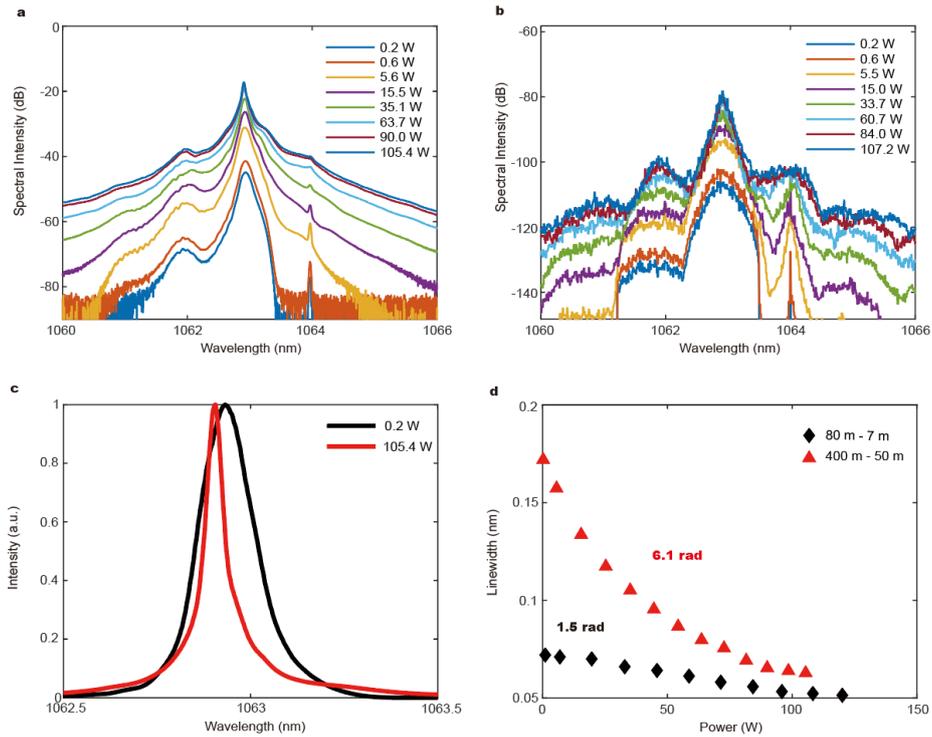

**Fig. 3 Spectral narrowing in the high-power fiber amplifier. a**, Measured optical spectra at increasing output power levels for the case of B-integral 6.1 rad. **b**, Corresponding simulated spectral evolution at increasing output power. **c**, Normalized output spectra at 0.2 W (low power) and 105.4 W (high power) for clear comparison. **d**, Linewidth variation as a function of output power for two configurations: a B-integral of 6.1 rad (achieved with 400 m nonlinear broadening fiber, and 50 m delivery fiber used) and 1.5 rad (using 80 m and 7 m fibers, respectively).

A notable observation is the emergence of a narrow peak at the FWM pump wavelength in both experiment and simulation. Simulations assumed 50 dB suppression at this wavelength, but the residual 1063.9 nm pump laser experiences amplification. Parametric gain provided by the first- and second-order FWM signals (at 1062.9 nm and 1061.9 nm, respectively) amplifies this component. Then, cross-phase modulation from co-propagating signals induces broadening of the 1063.9 nm line, which in turn contributes to more sideband formation. The result indicates effective filtering of the OPC output is critical to achieving optimal performance.

The central part of the output spectrum demonstrates clear spectral narrowing, though influenced by sidebands. Fig. 3c compares normalized spectra at 0.24 W and 105.4 W, showing significant linewidth reduction accompanied by a raised spectral pedestal. Linewidth dependence on output power is summarized in Fig. 3d, which includes an additional dataset obtained using an 80 m PM980 broadening fiber and a 7 m delivery fiber—corresponding to a B-integral of ~1.5 rad at 120 W output. For the high-nonlinearity case (6.1 rad), linewidth compressed from 0.172 nm to 0.063 nm, yet failed to recover the original seed linewidth of 0.049 nm. In contrast, complete compression from 0.072 nm to 0.049 nm was achieved under the low-nonlinearity condition (1.5 rad). This linewidth-dependent performance is attributed to phase mismatch in the fiber-based FWM process in OPC, which imposes a wavelength-dependent phase shift on the conjugated field. This shift becomes more pronounced for broader linewidths and cannot be compensated during amplification. These results underscore the necessity for optimized OPC generation in future work to achieve ideal spectral reconstruction.

Figure 4a presents simulated spectral evolution along the 50 m delivery fiber for a B-integral of 6.1 rad. The phase-conjugated laser spectrum undergoes progressive compression with propagation distance, with sidebands reaching minimal intensity around 25–30 m before increasing again upon further propagation. These features are attributed to non-ideal filtering prior to amplification. The residual pump laser of the OPC stage is regenerated within the first 10 m of fiber and undergoes linewidth broadening thereafter. The rate of spectral compression varies along the fiber: an initial rapid narrowing is followed by more gradual changes in linewidth, consistent with the trends observed in Fig. 3d. For comparison, Fig. 4b shows the spectral evolution without phase conjugation to the input, demonstrating rapid and monotonic broadening typical of conventional high-power fiber amplifiers. This contrast underscores the critical role of phase conjugation in enabling spectral compression.

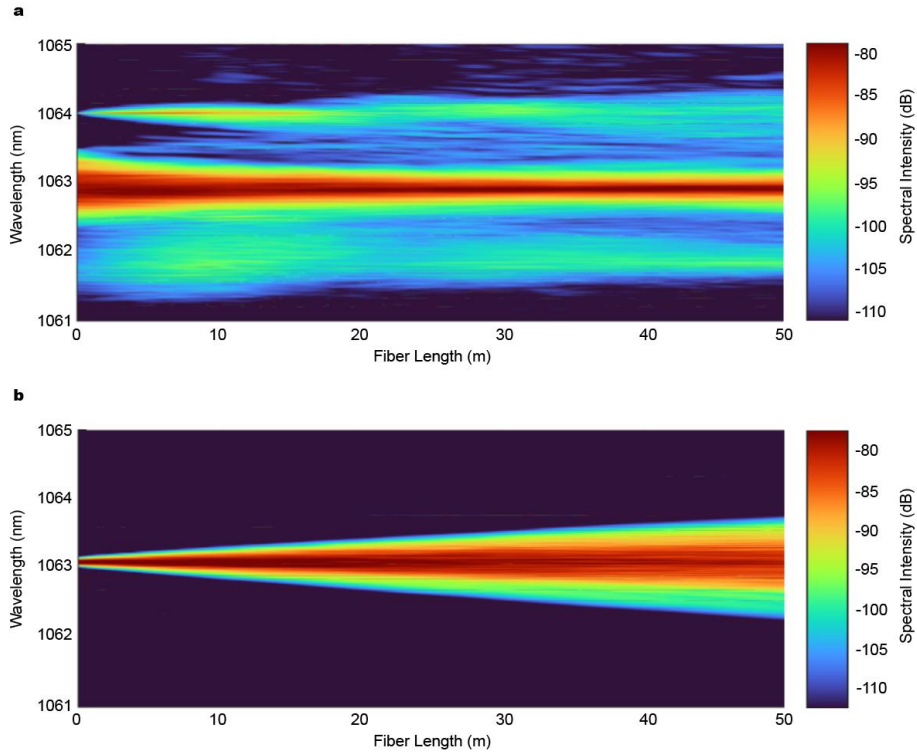

**Fig. 4 Simulated spectral evolution along the 50 m delivery fiber under a B-integral of 6.1 rad. a**, With injection of the phase-conjugated signal, as implemented in the experiment. **b**, Without phase conjugation, showing typical progressive spectral broadening.

A notable observation is the discrepancy between the linewidth of the conjugated signal and that of the pre-OPC broadened seed: 0.172 nm vs. 0.205 nm for the 6.1 rad case, and 0.072 nm vs. 0.087 nm for the 1.5 rad case. This reduction in effective linewidth is undesirable, as narrower starting linewidths in the amplifier provide less SBS suppression. Implementing dispersion management with extra propagating fiber before the amplifier increased the starting linewidth and improved final compression performance, as demonstrated in Supplementary Information Section 4, Fig. 4.1.

**SBS suppression**

The efficacy of SBS suppression with the phase conjugation approach was evaluated experimentally. Fig. 5 compares the SBS performance of phase-modulated and phase-conjugated systems under comparable initial linewidth conditions, using a 0.1% backward power level as the SBS threshold criterion. The phase-modulated system employed a single-frequency seed with 10 GHz electro-optic modulation bandwidth under white noise excitation. This system reached the SBS threshold at 69 W, with an initial linewidth of 0.044 nm and an output linewidth of 0.048 nm. In contrast, the phase-conjugated system started with a linewidth of 0.049 nm. At an output power of 105.4 W, the backward power remained at 26 mW—well below the SBS threshold. Due to the limited power capacity of the amplifier, the exact SBS threshold could not be measured directly. By extrapolating the backward power using an exponential model—appropriate for SBS in the small-signal regime—the phase conjugation approach is estimated to provide approximately 2.5 times higher SBS threshold compared to the phase-modulated system.

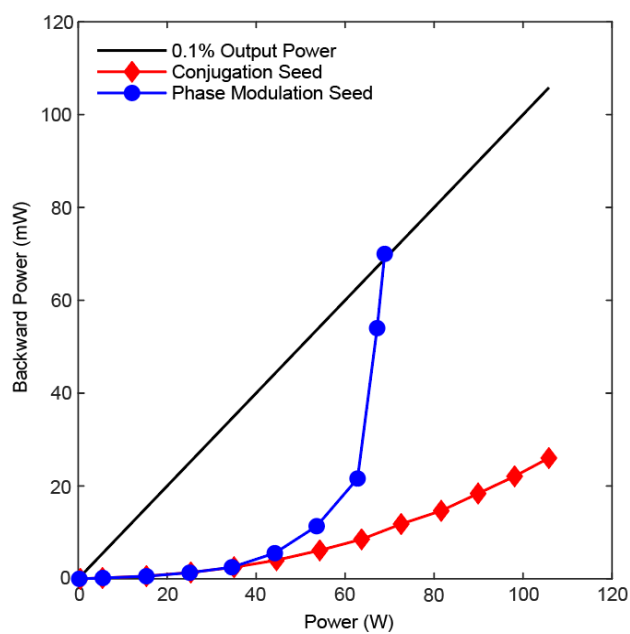

**Fig. 5 Comparison of SBS performance between phase-conjugated and phase-modulated laser systems.** Measured backward power as a function of output power for the phase-conjugated master oscillator configuration and a conventional phase-modulated seed laser with comparable initial linewidth. The black line indicates 0.1% of the output power, used to determine the SBS threshold.

The effectiveness of this method hinges on the interplay between spectral compression and SBS gain. A higher degree of spectral compression under equivalent nonlinearity is highly desirable. In this work, a linewidth reduction from 0.207 nm to 0.065 nm was achieved with a B-integral of 6.1 rad, whereas a co-modulation approach[33] required a B-integral of 18.7 rad to compress the linewidth from 0.2515 nm to 0.0838 nm. This implies that, for the same B-integral, the phase conjugation technique yields approximately three times greater spectral compression, corresponding to superior SBS suppression. The enhanced compression efficiency is attributed to the multi-longitudinal-mode nature of the FBG-based master oscillator used here, which facilitates more effective linewidth broadening and recompression.

With improvement in spectral compression and broadening rates, it may become possible to operate in a regime where the SBS threshold is effectively avoided in an optimized phase-conjugated master oscillator power amplifier system. In such a design, nonlinear phase accumulation must be carefully balanced between the broadening stage and the amplifier stage. Under these conditions, increases in either output power or delivery fiber length would lead to rapid spectral broadening rather than SBS. If the broadening occurs sufficiently quickly, the system would emit a broader linewidth without triggering SBS. This behavior is particularly advantageous for practical consideration, as it prevents catastrophic SBS-induced damage even under non-ideal operational settings.

**Demodulation of nonlinear phase inside the master oscillator**

Experiments conducted under low pre-OPC broadening conditions revealed a notable phenomenon of over-compression, in which the output linewidth becomes narrower than that of the original oscillator. As shown in Fig. 6, the linewidth evolution was measured in a configuration using an 80 m nonlinear broadening fiber and a 50 m delivery fiber. The initial oscillator linewidth was 0.049 nm, which broadened to 0.067 nm after the nonlinear pre-broadening and OPC stages. As the amplifier power increased, the linewidth was compressed to approximately 0.040 nm—narrower than the direct output from the oscillator. This observation suggests that not only is the nonlinear phase accumulated during external broadening compensated after OPC and amplification, but also the nonlinear phase originating within the FBG-based oscillator itself. This behavior is consistent with the fact that the physical mechanisms governing spectral broadening inside and outside the cavity are identical. These findings suggest a potential route to simplify the phase-conjugation approach by eliminating the dedicated external broadening stage, thereby reducing system complexity while retaining performance benefits.

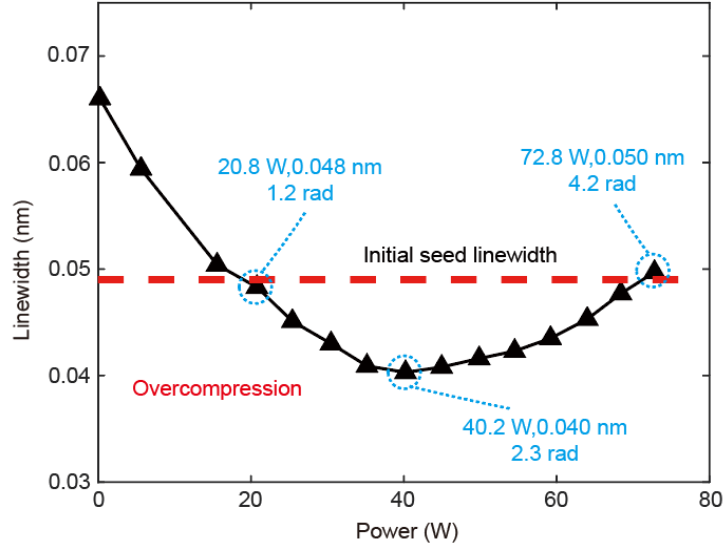

**Fig. 6 Demodulation of nonlinear phase inside the master oscillator.** Output linewidth as a function of output power for a system incorporating an 80 m nonlinear broadening fiber and a 50 m delivery fiber.

**Conclusion**

In this study, we have proposed and experimentally validated a novel all-optical approach for spectral linewidth compression and SBS suppression in high-power narrow-linewidth fiber amplifiers, based on optical phase conjugation. By leveraging nonlinear spectral broadening followed by phase conjugation, this method enables linewidth narrowing in fiber amplifier, resulting in narrow linewidth output and enhanced SBS threshold. Key results of a proof of principle experiment include a spectral compression ratio exceeding 3 times and an SBS threshold enhancement of about 2.5 times compared to conventional phase-modulated systems. Interestingly, under low pre-broadening conditions, the linewidth was compressed beyond that of the original oscillator, indicating that nonlinearities originating both inside and outside the oscillator cavity can be compensated.

All-optical nature of approach offers high bandwidth and operational flexibility absent in electro-optic modulation schemes. Employing an FBG-based fiber oscillator as the seed laser not only reduces system cost but also improves the spectral compression rate. Despite these advantages, limitations related to phase mismatch in fiber-based OPC and residual sidebands due to imperfect filtering require further investigation. Future work will focus on optimizing the OPC process and spectral filtering method, integrating dispersion-management strategies, and exploring higher-power demonstrations with improved spectral fidelity. This approach represents a paradigm shift in nonlinearity management and power scaling of narrow-linewidth fiber lasers, and holds significant potential for substantially enhancing the spectral brightness of current kilowatt-class fiber laser systems.

**Methods**

**Simulation details**

OPC and spectral compression processes presented in Fig. 1 were simulated by numerically solving the NLSE using the split-step Fourier method. The fiber in Fig. 1b is a 200 m passive fiber with a nonlinear coefficient $\gamma_1$ = 2.1 /W/km and a loss coefficient $\alpha_1$ = 1.3 dB/km. Dispersion properties at relevant wavelengths were derived from the Sellmeier equation of silica. The fiber in Fig. 1c is a 10 m active gain fiber with a gain coefficient of 4.8 dB/m, while maintaining the same nonlinear and dispersion parameters. The input signal was a continuous-wave laser with 4 W power and a linewidth of 15 GHz.

Simulations corresponding to Fig. 3 and Fig. 4 were also conducted using the NLSE and SSFM, with parameters consistent with the experimental configurations described below.

**Experimental setup**

The experimental configuration is illustrated in Fig. 2a. A master oscillator is constructed using a pair of FBGs centered at 1064.9 nm with a bandwidth of 0.045 nm, along with 2.5 m of ytterbium-doped fiber (YDF). This oscillator produces an output with a maximum linewidth of 0.049 nm. A pre-amplifier increases the power to regulate the total nonlinear phase accumulation, delivering up to 5.02 W of amplified output.

The seed is then passed through either an 80 m or 400 m PM980 fiber to accumulate nonlinear phase shift (B-integral). The nonlinear coefficient $\gamma$ of the PM980 fiber is calculated to be 3.66×10⁻³ /W/m, based on a nonlinear refractive index $n_2$ = 2.7×10⁻²⁰ m²/W and an effective mode-field area $A_{eff}$ = 42 μm² at the operating wavelength.

The broadened seed is combined with a single frequency pump laser at 1063.9 nm (maximum power 7.8 W) via a 50:50 fiber coupler. FWM occurs in a 20 m PM980 fiber, generating a phase-conjugated signal near 1062.9 nm. Two filter stages consisting of circulator and 1062 nm high-reflectivity FBG removes residual pump and original seed wavelengths, retaining only the phase-conjugated signal for final amplification.

The high-power amplifier comprises two stages: a pre-amplifier pumped by a 9 W, 976 nm laser through 3.5 m of PM-10/125-YDF, raising the signal power to ~2 W; and a main amplifier using a 200 W, 976 nm pump with 1.8 m of PM-15/130-YDF. To further increase the nonlinear phase accumulation, a delivery fiber (7 m or 50 m of PM-15/130-GDF) is spliced after the amplifier. The PM-15/130-GDF fiber has a mode-field area of 140 μm² and a nonlinear coefficient of ~$1.15\times10^{-3}$ /W/m. The output power reaches 105.4 W with the 50 m delivery fiber, corresponding to a B-integral of approximately 6.1 rad.

**Data availability**

Some data from this study are available in the article and the Supplementary Information. Further datasets are available from the corresponding author on reasonable request.

**Competing interests**

The authors declare no competing interests.

**Author contributions**

The research was initiated and led by Yan Feng. The experiment was planned and designed by Tingwei Gu and Yan Feng and performed by Tingwei Gu with contributions from Xin Zeng, Xuezong Yang, Huawei Jiang, Suming Luo, Maokai Yang. The data analysis was conducted by Tingwei Gu. Simulations were conducted by Tingwei Gu. Tingwei Gu and Yan Feng discussed the results and wrote the paper.

**Funding**

National Natural Science Foundation of China (U2541226, 12341404), Hangzhou Institute for Advanced Study (2023HIAS-V004).